%% file: QSTQPT-2.tex
\documentclass[prl,aps,twocolumn,superscriptaddress,amsmath,amssymb, showpacs]{revtex4}
\usepackage{epsfig}
\newcommand{\Tr}{\mathrm{Tr}}
\input{macros.tex}

\newcommand{\ket}[1]{| #1\rangle}
\newcommand{\bra}[1]{\langle #1|}
\newcommand{\beqa}{\begin{eqnarray}}
\newcommand{\eeqa}{\end{eqnarray}}
\newcommand{\beq}{\begin{equation}}
\newcommand{\eeq}{\end{equation}}

\begin{document}
\title{Selective and Efficient Quantum State Tomography and its Application to Quantum Process Tomography}

\author{Ariel Bendersky}
\affiliation{ICFO-Institut de Ciencies Fotoniques, Mediterranean Technology Park,
08860 Castelldefels (Barcelona), Spain}
\author{Juan Pablo Paz}
\affiliation{Departamento de F\'\i sica FCEyN UBA and IFIBA CONICET-UBA, Pabell\'on 1, Ciudad Universitaria, 1428 Buenos Aires, Argentina}
\date{\today}

\begin{abstract}
We present a method for quantum state tomography that enables the efficient estimation, with fixed precision, of any of the matrix elements of the density matrix of a state, provided that the states from the basis in which the matrix is written can be efficiently prepared in a controlled manner. Furthermore, we show how this algorithm is  well suited for quantum process tomography, enabling to perform selective and efficient quantum process tomography.
\end{abstract}

\maketitle

\section{Introduction}

Quantum information processing tasks always involve the preparation and manipulation of quantum systems. To be able to perform such tasks it is essential to be able to characterize both quantum states and quantum operations. The protocols for characterizing a quantum state are usually refered to as quantum state tomography (QST)\cite{NielsenChuang, QST, CompressedSensing, EQST, pazNature, paperConTerra}. In general, QST is a hard task since it involves an exponentially large number of measurements to be preformed (exponentially large on the number of subsystems). Not only that, but the type of measurements that one needs to perform on the systems are usually not easy to perform.

On the other hand, the characterization of quantum processes, known as quantum process tomography (QPT)\cite{NielsenChuang, AAPT, MohsRezLid, DCQD}, is also an exponentially hard task. However, there are some quantum algorithms that allow to efficiently extract important information about a given quantum process \cite{SCNQP, CeciLopez1, CeciLopez2, paper1conFer, paper2conFer, prlChris, paperConCeciLopez}. These algorithms do not require performing QST on the final states but measuring quantities such as survival probabilities (or transition probabilities). Other algorithms for QPT, however, do depend upon QST. This makes algorithms for QST an essential tool not just for state tomography, but for process tomography also.

In this paper we will present new methods for quantum state tomography. To be specific, it is useful to describe the quantum states we are to perform tomography in the following way:. Let $\mathcal H$ be the Hilbert space for the system in question, and let $\mathcal B=\left\lbrace \ket{\psi_a}, a=1,...,D \right\rbrace$ be a basis of $\mathcal H$, where $D=\text{dim} \mathcal H$. Then the density matrix $\rho$ of a state can be written in the basis $\mathcal B$ as
\begin{equation}\label{eqEstadoH}
 \rho=\sum_{a,b=1}^D \alpha_{ab}\ket{\psi_a}\bra{\psi_b}
\end{equation}
where $\alpha_{ab}=\bra{\psi_a}\rho\ket{\psi_b}$.

In what follows,  we will present a method for selective efficient quantum state tomography (SEQST) that allows to estimate any coefficient $\alpha_{ab}$ with resources scalling polynomially with the number of subsystems. For this to be possible, we need that any state from the basis $\mathcal B$ can be efficently prepared in a controlled manner. This method, when applied for implementing QPT results in a protocol for efficient and selective QPT that is equivalent to the one presented in \cite{paper1conFer, paper2conFer}, illustrating one of the virtues of such a selective and efficient QST scheme.

This paper if organized as follows. First we briefly review existing methods for QPT that rely on performing QST in the final states of a certain process. Then we present the selective and efficient algorithm for QST and show how it provides the right tool for efficient and selective QPT, as opposed to previous methods for QST. Finally, we compare that QPT algorithm to the one presented in \cite{paper1conFer, paper2conFer}, showing how both can be understood in a common theoretical frame.

\section{Quantum Process Tomography Based on State Tomography}

The goal of QPT is to identify the temporal evolution enforced by a certain quantum process. Any such process is mathematically represented by a linear map $\mathcal E$ transforming initial states into final states.  In fact, the operation $\mathcal E$ is not only linear but also completely positive, and acts as
\begin{equation}
 \mathcal E(\rho_{in})=\rho_{out}
\end{equation}
This operation represents the discrete (input--output) evolution of quantum states. We will focus on maps that are trace preserving and whose output dimension is the same as the input one. To describe the quantum map it is convenient to parametrize it in some way. It is simple to notice that any linear map can be written in terms of a certain matrix, known as the $\chi$--matrix. This is defined with respect to a certain basis of the space of operators. In fact, if we choose a basis $\left\lbrace E_m, m=0,...,D^2-1\right\rbrace$, the $\chi$--matrix representation for $\mathcal E$ is determined by the equation:
\begin{equation}\label{eq:ChiRepresentation}
\mathcal E(\rho) = \sum_{m n} \chi_{mn}E_m \rho E_{n}^\dagger.  
\end{equation}
This description is completely general since the above expressions can be written for any linear channel. Properties of the channel are in one to one correspondence with properties of the $\chi$--matrix. In fact, the channel preserves the hermiticity if and only if  the $\chi$--matrix is hermitian. Also, the channel preserves trace, if and only if the condition $\sum_{m n} \chi_{m n} E_{n}^\dagger E_m = I$ is satisfied. Fianlly, the channel is completely positive if and only if the $\chi$--matrix is positive. Thus, the $\chi$--matrix (which, as we mentioned above, depends on the choice of basis for the space of operators) fully describes the channel. Therefore, quantum process tomography is the task of estimating the matrix elements of $\chi$. To achieve this goal there are several methods, some of which involve performing quantum state tomography on final states. Let us review them now.  

\subsection{Ancilla Assisted Quantum Process Tomography}\label{secAAPT}

The Ancilla Assisted Quantum Process Tomography (AAPT) \cite{AAPT, MohsRezLid} uses $n$ ancilliary qubits and allows to extract all the information about the channel. However, as presented in \cite{AAPT, MohsRezLid}, it only allows to obtain full information about the process, being unable to only useful partial information about the channel. Thus it is inherently inefficient as full QPT always is. 

\begin{figure}[ht]
\begin{center}
\epsfig{file=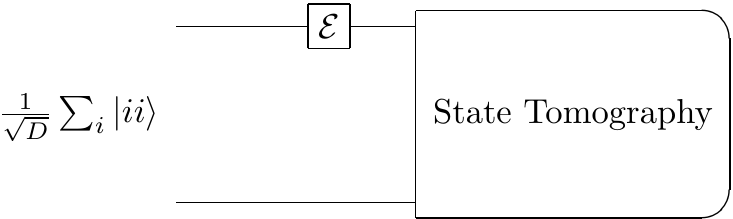, angle=0, width=0.45\textwidth}
\end{center}
\caption{Scheme for the Ancilla Assisted Process Tomography quantum algorithm.\label{figAAPT}}
\end{figure}

This is the first of two methods that we will mention here that exploit the state--channel duality given by Choi--Jamio\l{}kowski's isomorphism. Such isomorphism establishes a one to one relationship between linear operators from $\mathcal H \otimes \mathcal H$ to $\mathcal H \otimes \mathcal H$ and completely positive superoperators acting on the space of operators from $\mathcal H$ to $\mathcal H$. The Choi--Jamio\l{}kowski's isomorphisms establishes a correspondence between states and channels in the following way: 

\begin{equation}\label{eqCJ}
  \rho_\E = \mathcal{E}\otimes\mathbb I \left( \ket{I}\bra{I}\right).
\end{equation}
where $\ket{I}= \sum_i\ket{ii}/\sqrt{D}$ is the maximally entangled state. After the application of the channel to one of the parts, one can perform state tomography to the state $\rho_\E$, obtaining full information about the process $\E$. Figure \ref{figAAPT} ilustrates this algorithm.

One of the strenghts of  AAPT is that the initial state can be another state and not necessarily the maximally entangled state. As the number of independent parameters defining the initial state (the Shmidt number) is $D^2$, such state can always encode the necessary  information to define the quantum channel.

This method, besides requiring $n$ ancilliary qubits, has the following two troublesome properties: First, it is not clear if the information from the $\chi$ matrix can be directly accessed via measurements on the resulting states. Second but related to the previous point, it is not clear how to use QST on the final state to efficiently extract partial and relevant information on the channel. These two issues will be solved in what follows.

\subsection{Direct Characterization of Quantum Dynamics}\label{secDCQD}

The Direct Characterization of Quantum Dynamics (DQCD)\cite{MohsRezLid, DCQD} is a quantum algorithm similar to that of AAPT in many aspects. It also resorts to $n$ ancilliary qubits, and relies on the Choi--Jamio\l{}kowski isomorphism. Contrary to the AAPT, on the DCQD the authors explicitely showed a method to efficiently and selectively measure the diagonal coefficients of the $\chi$--matrix; however, off-diagonal elements still require to invert en exponentially large matrix. This makes the method inefficient. 

To describe the method let us consider the operator basis consisting of the $n$--fold tensor product Pauli operators acting on each qubit. We denote these operators as $\left\lbrace E_m, m=0,...,D^2-1 \right\rbrace$. In that basis, the channel description is given by
\begin{equation}
 \E\left(\rho\right)=\sum_{mn}\chi_{mn}E_m \rho E_n^\dagger
\end{equation}

In order to perform diagonal tomography, the algorithm proceeds as follows. First, as the AAPT, we have to generate the state that is isomorphic to the channel
\begin{equation}
  \rho_\E = \frac{1}{D}\sum_{ijmn}\chi_{mn}E_m\ket{i}\bra{j}E_n^\dagger\otimes\ket{i}\bra{j}.
\end{equation}
Now, the probability of measuring the state $\frac{1}{\sqrt{D}}\sum_i\ket{ii}$ on the output is given by
\begin{equation}
 \frac{1}{D}\sum_{kl}\bra{kk}\rho_\E\ket{ll}=\frac{1}{D^2}\sum_{mn}\chi_{mn}\Tr\left(E_m\right)\Tr\left(E_n^\dagger\right)=\chi_{00}.
\end{equation}
Thus, we see that the survival probability of the input state directly gives the coefficient $\chi_{00}$. It is easy to show that the very same method can be used to evaluate any diagonal coefficient of the $\chi$ matrix. In fact, the probability of obtaining the final state $\frac{1}{\sqrt{D}}\sum_iE_k\otimes\mathbb I\ket{ii}$ is nothing but $\chi_{kk}$. As the set 
\begin{equation}\label{eqBaseCJ}
 \mathcal R=\left\lbrace E_k\otimes\mathbb I\ket{I}, k=0,\ldots,D^2-1  \right\rbrace 
\end{equation}
forms an orthonormal basis of $\mathcal H\otimes\mathcal H$, a measurement in that basis will suffice for diagonal tomography of the $\chi$--matrix.

The main problem arises when off-diagonal tomography is taken into account. The solution presented in \cite{MohsRezLid, DCQD} is to use a state other than a maximally entangled one. However, it can be shown that in the most general case, this procedure requires inverting an exponentally large matrix. Again, this makes the method inefficient.

Below, we will introduce a method for QST that provides not only a convenient tool for QST, but also would provide the necessary ingredient missing in AAPT and DCQD to obtain an efficient and selective QPT.

\section{Selective and Efficient Quantum State Tomography}\label{secSEQST}

The standard method for QST was clearly described in \cite{NielsenChuang}. 
This method resorts to the description of the state in the Pauli operator basis as
\begin{equation}
 \rho=\frac{1}{D} \sum_i \text{Tr}\left(\rho E_i\right) E_i
\end{equation}
where $E_i$ are the $n$--fold Pauli operator basis for an $n$ qubit system. It is straightforward to perform tomography in this basis by just measuring the expectation value of every $E_i$. Although this method is indeed selective and efficient, it is not well suited for selective and efficient QPT. 

The method we are about to introduce, the Selective and Efficient Quantum State Tomography (SEQST), is also efficient and selective but, as opposed to the standard method, it is selective in any basis of the corresponding Hilbert space. That is, given the state written in the form
\begin{equation}
 \rho=\sum_{a,b=1}^D \alpha_{ab}\ket{\psi_a}\bra{\psi_b}
\end{equation}
and provided we know how to prepare the states from the corresponding basis in a controlled manner, we will be able to selectively measure any given coefficient $\alpha_{ab}$. Such measurement will be efficient, meaning that given a precision, the number of required single click measurements doesn't scale with the size of the system.

\begin{figure}[ht]
\begin{center}
\epsfig{file=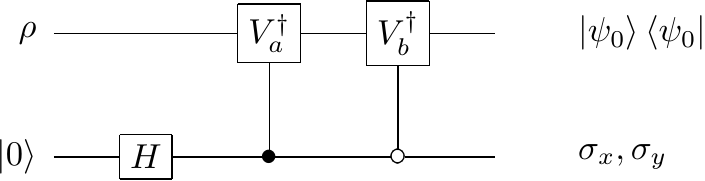, angle=0, width=0.45\textwidth}
\end{center}
\caption{Quantum circuit for Selective and Efficient Quantum State Tomography.\label{figTomEstNoDiagRed}}
\end{figure}

Consider the circuit shown in Fig. \ref{figTomEstNoDiagRed} where the operators $V_a$ are the ones that prepare the states from the basis $\mathcal B=\left\lbrace\ket{\psi_a}, a=1,...,D\right\rbrace$ from the state $\ket{\psi_0}$. That is, $V_a\ket{\psi_0}=\ket{\psi_a}$.

It is straightforward to verify that by measuring the average value of the operator $\ket{\psi_0}\bra{\psi_0}\otimes\sigma_x$ (that is the average value of the operator $\sigma_x$ conditioned on the detection of 
the state  $\ket{\psi_0}$ on the main system) one obtains the real part of $\chi_{ab}$. Moreover, replacing $\sigma_x$ by $\sigma_y$, the same method provides the imaginary part of the same matrix element. Thus, 
\begin{eqnarray}
 \Tr\left(\rho_F \ket{\psi_0}\bra{\psi_0}\otimes\sigma_x \right)&=&\Re{\chi_{ab}}\\
 \Tr\left(\rho_F \ket{\psi_0}\bra{\psi_0}\otimes\sigma_y \right)&=&\Im{\chi_{ab}}
\end{eqnarray}
where $\rho_F$ is the state prior to the measurement.

To discuss the efficiency of the method we should  analyze the resources needed by this algorithm. First of all, the efficiency of the method is limited by that of the implementation  of the controlled--$V_a^\dagger$ and controlled--$V_b^\dagger$ operators. In fact, if the implementation of such operations require $O\left(f\left(n\right)\right)$ operations, then the full circuit will also require $O\left(f\left(n\right)\right)$.

The other point to determine the efficiency of the method is to determine the number of experimental runs required to obtain the desired result to a given precision $\epsilon$ with a probability of success $p$. To answer that question we just need to consider that each experimental run gives one of three results ($+1$ corresponding to $\ket{\psi_0}$ on the system and $\ket{0}$ on the ancilla, $-1$ corresponding to $\ket{\psi_0}$ on the system and $\ket{1}$ on the ancilla, and $0$ corresponding to another result on the system). The $\chi$ matrix element is estimated by computing the average value these results after performing a certain number of repetitions $M$. Each of the results are detected at random with their corresponding probabilities. Therefore, one can use a Chernoff bound to show that to obtain the correct result with uncertainty $\epsilon$ and a probability $p$ of success, the number of experimental runs $M$ must be such that
\begin{equation}
 M\geq \frac{2 \ln\left(\frac{2}{p}\right) }{\epsilon^2}
\end{equation}
which does not depend on $n$ or $D$. This implies that the method is efficient.

\subsection{Application to Quantum Process Tomography}

In this section we will show how the SEQST algorithm is the right tool to perform selective and efficient QPT when combined with the AAPT method reviewed in Sec. \ref{secAAPT}.

In order to proceed, we need to find out the way in which the quantum state isomorphic to the channel $\E$ depends on the $\chi$--matrix of such channel. Thus, we will show that the $\chi$--matrix of the channel is nothing but the matrix element of the quantum state in a particular basis. Therefore, by performing quantum state tomography in that basis we directly provide the  information about the $\chi$--matrix of the channel. To show this, we use equation \eqref{eqCJ} and replace the expansion of $\E$ in the Pauli basis:

\begin{equation}
 \rho_\E = \sum_{mn} \chi_{mn} E_m\otimes\mathbb I  \ket{I}\bra{I} E_n^\dagger\otimes\mathbb I.
\end{equation}
Indeed, this shows that $\chi$ is the matrix  representation of $\rho_\E$ in the basis $\mathcal R$ shown in equation \eqref{eqBaseCJ}. Therefore, to selectively measure a single $\chi$ coefficient one only needs to perform selective tomography in the basis $\mathcal R$.

\begin{figure}[ht]
\epsfig{file=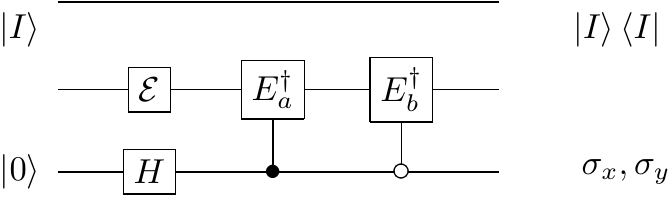, angle=0, width=0.45\textwidth}
\caption{Application of SEQST to QPT.\label{figSEQSTQPT}}
\end{figure}

Using the results already presented in the previous section, we see that to do this we should implement the circuit described in Figure \ref{figSEQSTQPT}. Here, the application of the channel $\E$ to one of the pieces of the maximally entangled state can be regarded as the preparation of the state isomoprhic to the channel. In turn, the rest of the circuit is nothing but the SEQST algorithm described above.

It is important to point out that, since the measurement is direct, the analysis of the resources required to implement the method (presented in the previous section) directly applies to this case. The only extra resources needed in this case are involved in the preparation and measurement of the maximally entangled state which require $O\left(n\right)$ aditional single and two qubit gates.

\section{Comparison with Other Selective and Efficient QPT Schemes}

\begin{figure}[ht]
\epsfig{file=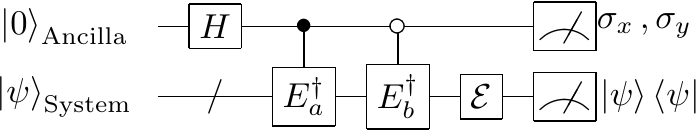, angle=0, width=0.45\textwidth}
\caption{Circuit for the Selective and Efficient Quantum Process Tomography algorithm. Depending on the measurement of $\sigma_x$ or $\sigma_y$, the real or imaginary part of $\chi_{ab}$ will be obtained.}
\label{circ:offdiag} 
\end{figure}

Another quantum algorithm for selective and efficient quantum process tomography is the one known as SEQPT, precisely for Selective and Efficient Quantum Process Tomography\cite{paper1conFer, paper2conFer}. The main idea there is to follow the proceedure described by the circuit shown in Figure \ref{circ:offdiag}, and to estimate the average answer averaging over the entire Hilbert space of the system using the Haar measure for that purpose. As it is shown in \cite{paper1conFer, paper2conFer}, that average cn be directly related to the matrix element $\chi_{ab}$ as
\begin{eqnarray}\label{eqReImChi}
 \int \left<\sigma_x\otimes \ket{\psi}\bra{\psi} \right>  d\psi &=& \frac{D\text{Re}\left(\chi_{ab} \right) +\delta_{ab}}{D+1}\\
\int \left<\sigma_y\otimes \ket{\psi}\bra{\psi} \right>  d\psi &=& \frac{D\text{Im}\left(\chi_{ab} \right)}{D+1}.
\end{eqnarray}
Moreover, it can be shown that the average over the entire Hilbert space can be efficiently estimated by randomly sampling over a special (and finite) set of states which is known as a $2$--designs. For these reasons, the method SEQPT is not only selective but also efficient. 

As we mentioned above, in the SEQPT scheme, the average is estimated by randomly sampling states. In the scheme we proposed above, the average is obtained  \emph{automatically} by the quantum correlation between both parts of the maximally entangled state, without the need to resort to randomly preparing and detecting the special states of the $2$--design.

\section{Summary}

In this paper we presented a novel quantum algorithm to perform selective and efficient quantum state tomography in any Hilbert space basis, given that the states from that basis can be efficiently prepared in a controlled manner. 

We then showed that, when properly combined with the Ancilla Assisted Process Tomography, it yields a protocol for QPT that is both selective and efficient. Finally, we showed that this protocol shares some properties with SEQPT, a method presented in \cite{paper1conFer, paper2conFer}. The main difference is that the use of ancillas is a way to avoid the preparation of the special states of the $2$--design and sampling on them. 

This work was partially supported by funds from CONICET, ANPCyT, UBACyT, EU Q-Essence and Spanish FIS2010-14830 projects. A.B. is funded by an ERC Starting Grant PERCENT.


%

\end{document}

%% file: macros.tex
\usepackage{amsmath}
\usepackage{bbm,amsfonts}

\def\E{{\mathcal E}}

\usepackage{amsthm}
\theoremstyle{definition}

%% file: QSTQPT-2.bbl
\begin{thebibliography}{16}
\expandafter\ifx\csname natexlab\endcsname\relax\def\natexlab#1{#1}\fi
\expandafter\ifx\csname bibnamefont\endcsname\relax
  \def\bibnamefont#1{#1}\fi
\expandafter\ifx\csname bibfnamefont\endcsname\relax
  \def\bibfnamefont#1{#1}\fi
\expandafter\ifx\csname citenamefont\endcsname\relax
  \def\citenamefont#1{#1}\fi
\expandafter\ifx\csname url\endcsname\relax
  \def\url#1{\texttt{#1}}\fi
\expandafter\ifx\csname urlprefix\endcsname\relax\def\urlprefix{URL }\fi
\providecommand{\bibinfo}[2]{#2}
\providecommand{\Eprint}[2][]{\url{#2}}

\bibitem[{\citenamefont{Nielsen and Chuang}(2000)}]{NielsenChuang}
\bibinfo{author}{\bibfnamefont{M.~A.} \bibnamefont{Nielsen}} \bibnamefont{and}
  \bibinfo{author}{\bibfnamefont{I.~L.} \bibnamefont{Chuang}},
  \emph{\bibinfo{title}{Quantum Computation and Quantum Information}}
  (\bibinfo{publisher}{Cambridge University Press}, \bibinfo{year}{2000}).




\bibitem [{\citenamefont {Paris}\ and\ \citenamefont
  {{\v{R}}eh{\'a}{\v{c}}ek}(2004)}]{QST}%
  {\bibinfo {author} {\bibfnamefont {M.}~\bibnamefont
  {Paris}}\ and\ \bibinfo {author} {\bibfnamefont {J.}~\bibnamefont
  {{\v{R}}eh{\'a}{\v{c}}ek}},\ }{\emph {\bibinfo {title}
  {Quantum state estimation}}},\ Lecture notes in physics\ (\bibinfo
  {publisher} {Springer},\ \bibinfo {year} {2004})%
\bibitem [{\citenamefont {Gross}\ \emph {et~al.}(2010)\citenamefont {Gross},
  \citenamefont {Liu}, \citenamefont {Flammia}, \citenamefont {Becker},\ and\
  \citenamefont {Eisert}}]{CompressedSensing}%
  \bibfield  {author} {\bibinfo {author} {\bibfnamefont {D.}~\bibnamefont
  {Gross}}, \bibinfo {author} {\bibfnamefont {Y.-K.}\ \bibnamefont {Liu}},
  \bibinfo {author} {\bibfnamefont {S.~T.}\ \bibnamefont {Flammia}}, \bibinfo
  {author} {\bibfnamefont {S.}~\bibnamefont {Becker}}, \ and\ \bibinfo {author}
  {\bibfnamefont {J.}~\bibnamefont {Eisert}},\ } {\bibfield  {journal} {\bibinfo  {journal}
  {Phys. Rev. Lett.}\ }\textbf {\bibinfo {volume} {105}},\ \bibinfo {pages}
  {150401} (\bibinfo {year} {2010})}%
\bibitem [{\citenamefont {Cramer}\ \emph {et~al.}(2010)\citenamefont {Cramer},
  \citenamefont {Plenio}, \citenamefont {Flammia}, \citenamefont {Somma},
  \citenamefont {Gross}, \citenamefont {Bartlett}, \citenamefont
  {Landon-Cardinal}, \citenamefont {Poulin},\ and\ \citenamefont {Liu}}]{EQST}%
  \bibfield  {author} {\bibinfo {author} {\bibfnamefont {M.}~\bibnamefont
  {Cramer}}, \bibinfo {author} {\bibfnamefont {M.}~\bibnamefont {Plenio}},
  \bibinfo {author} {\bibfnamefont {S.}~\bibnamefont {Flammia}}, \bibinfo
  {author} {\bibfnamefont {R.}~\bibnamefont {Somma}}, \bibinfo {author}
  {\bibfnamefont {D.}~\bibnamefont {Gross}}, \bibinfo {author} {\bibfnamefont
  {S.}~\bibnamefont {Bartlett}}, \bibinfo {author} {\bibfnamefont
  {O.}~\bibnamefont {Landon-Cardinal}}, \bibinfo {author} {\bibfnamefont
  {D.}~\bibnamefont {Poulin}}, \ and\ \bibinfo {author} {\bibfnamefont {Y.-K.}\
  \bibnamefont {Liu}},\ } {\bibfield  {journal} {\bibinfo
  {journal} {Nature Comm.}\ }\textbf {\bibinfo {volume} {1}},\ \bibinfo {pages}
  {149} (\bibinfo {year} {2010})},\ \Eprint {http://arxiv.org/abs/1101.4366}
  {1101.4366}
\bibitem [{\citenamefont {Miquel}\ \emph {et~al.}(2002)\citenamefont {Miquel},
  \citenamefont {Paz}, \citenamefont {Saraceno}, \citenamefont {Knill},
  \citenamefont {Laflamme},\ and\ \citenamefont {Negrevergne}}]{pazNature}%
  \bibfield  {author} {\bibinfo {author} {\bibfnamefont {C.}~\bibnamefont
  {Miquel}}, \bibinfo {author} {\bibfnamefont {J.~P.}\ \bibnamefont {Paz}},
  \bibinfo {author} {\bibfnamefont {M.}~\bibnamefont {Saraceno}}, \bibinfo
  {author} {\bibfnamefont {E.}~\bibnamefont {Knill}}, \bibinfo {author}
  {\bibfnamefont {R.}~\bibnamefont {Laflamme}}, \ and\ \bibinfo {author}
  {\bibfnamefont {C.}~\bibnamefont {Negrevergne}},\ } {\bibfield  {journal} {\bibinfo  {journal} {Nature}\
  }\textbf {\bibinfo {volume} {418}},\ \bibinfo {pages} {59} (\bibinfo {year}
  {2002})}
\bibitem [{\citenamefont {Bendersky}\ \emph
  {et~al.}(2009{\natexlab{a}})\citenamefont {Bendersky}, \citenamefont {Paz},\
  and\ \citenamefont {Cunha}}]{paperConTerra}%
  \bibfield  {author} {\bibinfo {author} {\bibfnamefont {A.}~\bibnamefont
  {Bendersky}}, \bibinfo {author} {\bibfnamefont {J.~P.}\ \bibnamefont {Paz}},
  \ and\ \bibinfo {author} {\bibfnamefont {M.~T.}\ \bibnamefont {Cunha}},\
  } {\bibfield  {journal}
  {\bibinfo  {journal} {Phys. Rev. Lett.}\ }\textbf {\bibinfo {volume} {103}},\
  \bibinfo {pages} {040404} (\bibinfo {year} {2009}{\natexlab{a}})}
\bibitem [{\citenamefont {Altepeter}\ \emph {et~al.}(2003)\citenamefont
  {Altepeter}, \citenamefont {Branning}, \citenamefont {Jeffrey}, \citenamefont
  {Wei}, \citenamefont {Kwiat}, \citenamefont {Thew}, \citenamefont {O'Brien},
  \citenamefont {Nielsen},\ and\ \citenamefont {White}}]{AAPT}%
  \bibfield  {author} {\bibinfo {author} {\bibfnamefont {J.~B.}\ \bibnamefont
  {Altepeter}}, \bibinfo {author} {\bibfnamefont {D.}~\bibnamefont {Branning}},
  \bibinfo {author} {\bibfnamefont {E.}~\bibnamefont {Jeffrey}}, \bibinfo
  {author} {\bibfnamefont {T.~C.}\ \bibnamefont {Wei}}, \bibinfo {author}
  {\bibfnamefont {P.~G.}\ \bibnamefont {Kwiat}}, \bibinfo {author}
  {\bibfnamefont {R.~T.}\ \bibnamefont {Thew}}, \bibinfo {author}
  {\bibfnamefont {J.~L.}\ \bibnamefont {O'Brien}}, \bibinfo {author}
  {\bibfnamefont {M.~A.}\ \bibnamefont {Nielsen}}, \ and\ \bibinfo {author}
  {\bibfnamefont {A.~G.}\ \bibnamefont {White}},\ } {\bibfield  {journal} {\bibinfo  {journal}
  {Phys. Rev. Lett.}\ }\textbf {\bibinfo {volume} {90}},\ \bibinfo {pages}
  {193601} (\bibinfo {year} {2003})}
\bibitem [{\citenamefont {Mohseni}\ \emph {et~al.}(2008)\citenamefont
  {Mohseni}, \citenamefont {Rezakhani},\ and\ \citenamefont
  {Lidar}}]{MohsRezLid}%
  \bibfield  {author} {\bibinfo {author} {\bibfnamefont {M.}~\bibnamefont
  {Mohseni}}, \bibinfo {author} {\bibfnamefont {A.~T.}\ \bibnamefont
  {Rezakhani}}, \ and\ \bibinfo {author} {\bibfnamefont {D.~A.}\ \bibnamefont
  {Lidar}},\ }{\bibfield  {journal}
  {\bibinfo  {journal} {Phys. Rev. A}\ }\textbf {\bibinfo {volume} {77}},\
  \bibinfo {pages} {032322} (\bibinfo {year} {2008})}
\bibitem [{\citenamefont {Mohseni}\ and\ \citenamefont {Lidar}(2006)}]{DCQD}%
  \bibfield  {author} {\bibinfo {author} {\bibfnamefont {M.}~\bibnamefont
  {Mohseni}}\ and\ \bibinfo {author} {\bibfnamefont {D.~A.}\ \bibnamefont
  {Lidar}},\ }{\bibfield
  {journal} {\bibinfo  {journal} {Phys. Rev. Lett.}\ }\textbf {\bibinfo
  {volume} {97}},\ \bibinfo {pages} {170501} (\bibinfo {year}
  {2006})}
\bibitem [{\citenamefont {Emerson}\ \emph {et~al.}(2007)\citenamefont
  {Emerson}, \citenamefont {Silva}, \citenamefont {Moussa}, \citenamefont
  {Ryan}, \citenamefont {Laforest}, \citenamefont {Baugh}, \citenamefont
  {Cory},\ and\ \citenamefont {Laflamme}}]{SCNQP}%
  \bibfield  {author} {\bibinfo {author} {\bibfnamefont {J.}~\bibnamefont
  {Emerson}}, \bibinfo {author} {\bibfnamefont {M.}~\bibnamefont {Silva}},
  \bibinfo {author} {\bibfnamefont {O.}~\bibnamefont {Moussa}}, \bibinfo
  {author} {\bibfnamefont {C.}~\bibnamefont {Ryan}}, \bibinfo {author}
  {\bibfnamefont {M.}~\bibnamefont {Laforest}}, \bibinfo {author}
  {\bibfnamefont {J.}~\bibnamefont {Baugh}}, \bibinfo {author} {\bibfnamefont
  {D.~G.}\ \bibnamefont {Cory}}, \ and\ \bibinfo {author} {\bibfnamefont
  {R.}~\bibnamefont {Laflamme}},\ }
  {\bibfield  {journal} {\bibinfo  {journal} {Science}\ }\textbf {\bibinfo
  {volume} {317}},\ \bibinfo {pages} {1893} (\bibinfo {year}
  {2007})}
\bibitem [{\citenamefont {L\'opez}\ \emph {et~al.}(2009)\citenamefont
  {L\'opez}, \citenamefont {L\'evi},\ and\ \citenamefont {Cory}}]{CeciLopez1}%
  \bibfield  {author} {\bibinfo {author} {\bibfnamefont {C.~C.}\ \bibnamefont
  {L\'opez}}, \bibinfo {author} {\bibfnamefont {B.}~\bibnamefont {L\'evi}}, \
  and\ \bibinfo {author} {\bibfnamefont {D.~G.}\ \bibnamefont {Cory}},\ }{\bibfield  {journal} {\bibinfo
  {journal} {Phys. Rev. A}\ }\textbf {\bibinfo {volume} {79}},\ \bibinfo
  {pages} {042328} (\bibinfo {year} {2009})}
\bibitem [{\citenamefont {L\'evi}\ \emph {et~al.}(2007)\citenamefont {L\'evi},
  \citenamefont {L\'opez}, \citenamefont {Emerson},\ and\ \citenamefont
  {Cory}}]{CeciLopez2}%
  \bibfield  {author} {\bibinfo {author} {\bibfnamefont {B.}~\bibnamefont
  {L\'evi}}, \bibinfo {author} {\bibfnamefont {C.~C.}\ \bibnamefont {L\'opez}},
  \bibinfo {author} {\bibfnamefont {J.}~\bibnamefont {Emerson}}, \ and\
  \bibinfo {author} {\bibfnamefont {D.~G.}\ \bibnamefont {Cory}},\ }{\bibfield  {journal} {\bibinfo
  {journal} {Phys. Rev. A}\ }\textbf {\bibinfo {volume} {75}},\ \bibinfo
  {pages} {022314} (\bibinfo {year} {2007})}
\bibitem [{\citenamefont {Bendersky}\ \emph {et~al.}(2008)\citenamefont
  {Bendersky}, \citenamefont {Pastawski},\ and\ \citenamefont
  {Paz}}]{paper1conFer}%
  \bibfield  {author} {\bibinfo {author} {\bibfnamefont {A.}~\bibnamefont
  {Bendersky}}, \bibinfo {author} {\bibfnamefont {F.}~\bibnamefont
  {Pastawski}}, \ and\ \bibinfo {author} {\bibfnamefont {J.~P.}\ \bibnamefont
  {Paz}},\ }{\bibfield
  {journal} {\bibinfo  {journal} {Phys. Rev. Lett.}\ }\textbf {\bibinfo
  {volume} {100}},\ \bibinfo {pages} {190403} (\bibinfo {year}
  {2008})}
\bibitem [{\citenamefont {Bendersky}\ \emph
  {et~al.}(2009{\natexlab{b}})\citenamefont {Bendersky}, \citenamefont
  {Pastawski},\ and\ \citenamefont {Paz}}]{paper2conFer}%
  \bibfield  {author} {\bibinfo {author} {\bibfnamefont {A.}~\bibnamefont
  {Bendersky}}, \bibinfo {author} {\bibfnamefont {F.}~\bibnamefont
  {Pastawski}}, \ and\ \bibinfo {author} {\bibfnamefont {J.~P.}\ \bibnamefont
  {Paz}},\ }{\bibfield  {journal}
  {\bibinfo  {journal} {Phys. Rev. A}\ }\textbf {\bibinfo {volume} {80}},\
  \bibinfo {pages} {032116} (\bibinfo {year} {2009}{\natexlab{b}})}
\bibitem [{\citenamefont {Schmiegelow}\ \emph {et~al.}(2011)\citenamefont
  {Schmiegelow}, \citenamefont {Bendersky}, \citenamefont {Larotonda},\ and\
  \citenamefont {Paz}}]{prlChris}%
  \bibfield  {author} {\bibinfo {author} {\bibfnamefont {C.~T.}\ \bibnamefont
  {Schmiegelow}}, \bibinfo {author} {\bibfnamefont {A.}~\bibnamefont
  {Bendersky}}, \bibinfo {author} {\bibfnamefont {M.~A.}\ \bibnamefont
  {Larotonda}}, \ and\ \bibinfo {author} {\bibfnamefont {J.~P.}\ \bibnamefont
  {Paz}},\ }{\bibfield
  {journal} {\bibinfo  {journal} {Phys. Rev. Lett.}\ }\textbf {\bibinfo
  {volume} {107}},\ \bibinfo {pages} {100502} (\bibinfo {year}
  {2011})}
\bibitem [{\citenamefont {L\'opez}\ \emph {et~al.}(2010)\citenamefont
  {L\'opez}, \citenamefont {Bendersky}, \citenamefont {Paz},\ and\
  \citenamefont {Cory}}]{paperConCeciLopez}%
  \bibfield  {author} {\bibinfo {author} {\bibfnamefont {C.~C.}\ \bibnamefont
  {L\'opez}}, \bibinfo {author} {\bibfnamefont {A.}~\bibnamefont {Bendersky}},
  \bibinfo {author} {\bibfnamefont {J.~P.}\ \bibnamefont {Paz}}, \ and\
  \bibinfo {author} {\bibfnamefont {D.~G.}\ \bibnamefont {Cory}},\ }{\bibfield  {journal} {\bibinfo
  {journal} {Phys. Rev. A}\ }\textbf {\bibinfo {volume} {81}},\ \bibinfo
  {pages} {062113} (\bibinfo {year} {2010})}
\end{thebibliography}
